\documentclass{PoS}

\newcommand{\be}{\begin{equation}}
\newcommand{\ee}{\end{equation}}
\newcommand{\beq}{\begin{eqnarray}}
\newcommand{\eeq}{\end{eqnarry}}

\newcommand{\C}{{\kern+.25em\sf{C}\kern-.45em\sf{{\small{I}}}\kern+.45em\kern-.25em}}
\newcommand{\R}{{\kern+.25em\sf{R}\kern-.78em\sf{I} \kern+.78em\kern-.25em}}

\def\lsi{\raise0.3ex\hbox{$<$\kern-0.75em\raise-1.1ex\hbox{$\sim$}}}
\def\gsi{\raise0.3ex\hbox{$>$\kern-0.75em\raise-1.1ex\hbox{$\sim$}}}

\newcommand{\gsim}{\mathop{\gsi}}

\title{Spectral study of a chiral limit without chiral condensate}

\ShortTitle{Chiral limit without chiral condensate}

\author{\speaker{Wolfgang Bietenholz} \\ 
        Instituto de Ciencias Nucleares (ICN) \\
        Universidad Nacional Aut\'{o}noma de M\'{e}xico (UNAM) \\
        A.P. 70-543, C.P. 04510 Distrito Federal, M\'{e}xico \\
        E-mail: \email{wolbi@nucleares.unam.mx}}

\author{Ivan Hip \thanks{Work supported by the Croatian Ministry of
        Science, Education and Sports (project No.\ 0160013). \newline
        Most simulations for this project were
        performed on clusters of the ``Norddeutscher Verbund
        f\"{u}r Hoch- und H\"{o}chst\-leistungsrechnen'' (HLRN).
        We thank Hinnerk St\"{u}ben for technical assistance.} \\
        Faculty of Geotechnical Engineering \\
        University of Zagreb \\
        Hallerova aleja 7 \\
        42000 Vara\v{z}din, Croatia \\
        E-mail: \email{ivan.hip@gmail.com}}

\abstract{Random Matrix Theory (RMT) has elaborated successful 
predictions for Dirac spectra in field theoretical models. However,
a generic assumption by RMT has been a non-vanishing chiral
condensate $\Sigma$ in the chiral limit.
Here we consider the 2-flavour Schwinger model, where
this assumption does not hold. We simulated this model with
dynamical overlap hypercube fermions, and entered {\it terra incognita}
by analysing this Dirac spectrum. The usual RMT prediction
for the unfolded level spacing distribution in a unitary
ensemble is confirmed to a high precision.
The microscopic spectrum does not perform a Banks-Casher plateau.
Instead the obvious expectation is a density of the lowest 
eigenvalue $\lambda_{1}$ 
which increases $\propto \lambda_{1}^{1/3}$. That would correspond 
to a scale-invariant parameter $\propto \lambda V^{3/4}$, which is,
however, incompatible with our data. Instead we observe 
to high precision a scale-invariant parameter $z \propto \lambda V^{5/8}$.
This surprising result implies a microscopic spectral density
$\propto \lambda_1^{3/5}$, which still remains to be understood
in the light of RMT.}

\FullConference{The XXVII International Symposium on Lattice Field Theory - LAT2009\\
		 July 26-31 2009\\
		 Peking University, Beijing, China}

\begin{document}

\section{Dirac spectra and Random Matrix Theory}

Random Matrix Theory (RMT) has been applied extensively to predict
the microscopic spectral densities of Dirac operators in fermionic
quantum field theories. These predictions were well
confirmed, first with staggered fermions restricted to the 
sector of topological charge $\nu =0$ \cite{stag}, 
and later with Ginsparg-Wilson fermions also for charged 
topological sectors \cite{BJS,GWRMT}. These 
applications referred to the case of a {\em finite chiral condensate} 
$\Sigma = - \langle \bar \psi \psi \rangle$ in the chiral limit
of fermion mass $m \to 0$, which may occur spontaneously (as in QCD), 
or due to an anomaly (as in the 1-flavour Schwinger model). 
In this case the microscopic spectrum displays a plateau near zero;
its value is directly related to $\Sigma$ by the Banks-Casher formula
(in finite volume the plateau is slightly shifted away 
from zero). In the $\epsilon$-regime RMT predicts in addition a 
wiggle structure superimposed on this plateau, which does in 
fact agree with numerical data. Matching them to the RMT curves 
determines $\Sigma$ as the only free parameter --- this is a neat way 
to evaluate $\Sigma$.

There are also models with $\Sigma =0$, but that situation
is hardly explored by RMT. It occurs for instance in systems of
fermions interacting through Yang-Mills gauge theory above the
critical temperature of the chiral phase transition. There are 
numerical studies and conjectures about it \cite{airyref,highT}, 
but the features of such spectra remain controversial.

In addition there are models with $\Sigma (m \to 0)= 0$ even at zero
temperature. This is the case for the {\em 2-flavour Schwinger model,}
which we are going to discuss here. We simulated this model with
{\em dynamical Ginsparg-Wilson fermions} \cite{procprep}. Here we
present our observations on its Dirac spectrum,
which has been completely unexplored so far.

First we review the model and the analytical predictions for
$\Sigma$. Next we sketch our lattice formulation and
its simulation. Then we present the {\em unfolded
level spacing distribution}, which agrees accurately with the
generic RMT prediction for the unitary ensemble. As the main
subject of this report, we then focus on the probability 
density of the leading non-zero Dirac eigenvalue $\lambda_{1}$, 
where we reveal a surprising result. 
The {\em scale-invariant parameter} --- a rescaled eigenvalue
in finite volume --- does not agree with the obvious conjecture,
which is based on the critical exponent $\delta$ derived in
the literature. The microscopic spectrum does increase without a 
plateau, but its slope follows an {\em unexpected power law.}
Finally we also consider the bulk eigenvalues.

\section{The chiral condensate in the Schwinger model}

The Schwinger model corresponds to QED in $d=2$, given by the 
Lagrangian
\be
{\cal L} (\bar \Psi , \Psi , A_{\mu}) =
\bar \Psi (x) \left[ \gamma_{\mu} ( {\rm i} 
\partial_{\mu} + g A_{\mu} ) + m \right]     
\Psi (x) + \frac{1}{2} F_{\mu \nu}(x) F_{\mu \nu}(x) \ .
\ee
For $N_{f}$ degenerated fermion flavours of mass $m \ll g$,
a bosonised form of the Schwinger model leads to the
prediction \cite{Smi92},
\be
\Sigma (m) \propto  m^{1 / \delta} \ , 
\qquad \delta = \frac{N_{f}+1}{N_{f}-1} \ . 
\ee
\begin{itemize}

\item In the quenched case (formally $N_{f}=0$), the divergence of
$\Sigma (m \to 0)$ agrees with simulation results \cite{DHNS}.

\item For $N_{f}=1$ --- the original version of the Schwinger model ---
one obtains a finite value \
$\Sigma (m \to 0) = e^{\gamma} / ( 2 \pi^{3/2})$ \ due to the axial 
anomaly ($\, \gamma \,$ is Euler's constant) \cite{CJS}.

\item The explicit prediction for $N_{f}=2$ reads 
$\, \Sigma (m) \simeq 0.38 \, m^{1/3}$ \cite{Smilga}. \\
A numerical study with Domain Wall Fermions on a $16^{2}\times 6$ 
lattice measured $\Sigma$ in the range $m = 0.1 \dots 0.3$, and
a fit in this regime suggested
$\Sigma (m) \propto m^{0.388(68)}$ \cite{Hide}.

\end{itemize}

\section{Lattice formulation and simulation of the 2-flavour Schwinger model}

In our study, we use a lattice formulation with compact link variables
$U_{\mu, x} \in U(1)$ and the plaquette gauge action.
For the fermions we apply the {\em overlap hypercube Dirac operator,}
\be
D_{\rm ovHF}(m) = \Big( 1 - \frac{m}{2} \Big) \,
D_{\rm ovHF}^{(0)} + m  \ , \quad 
D_{\rm ovHF}^{(0)} = 1 + (D_{\rm HF} -1) /
\sqrt{(D_{\rm HF}^{\dagger} -1)( D_{\rm HF}-1) } \ .
\label{Dov}
\ee
$D_{\rm HF}(U)$ is a hypercube fermion operator \cite{WBIH}: it is
truncated perfect, and thus by construction approximately chiral 
\cite{WBEPJC}. In eq.\ (\ref{Dov}) it is inserted into the overlap
formula \cite{Neu}, which restores exact (lattice modified) 
chirality \cite{ML}. The spectrum of $D_{\rm ovHF}^{(0)}$ is located
on a unit circle in the complex plane, with centre 1. 
Compared to the standard overlap fermion formulation --- 
where the Wilson operator is inserted into the kernel
--- $D_{\rm ovHF}$  has a better level of locality, it 
approximates rotation symmetry better, and it has an improved 
scaling behaviour \cite{procprep,WBIH,WBEPJC,DovHF}. All these 
virtues are based on the similarity between the kernel and its 
overlap operator, $D_{\rm ovHF}^{(0)} \approx D_{\rm HF}$. 

In addition, that property also allows us to use a simplified
form of the HMC force term, given by a low polynomial in
$D_{\rm HF}$. This reduces the computational effort for dynamical
overlap fermions. The algorithm is kept exact by applying
$D_{\rm ovHF}$ to high precision in the Metropolis step at 
the end of each trajectory.

\begin{figure}[h!]
\begin{center}
\includegraphics[angle=270,width=.45\linewidth]{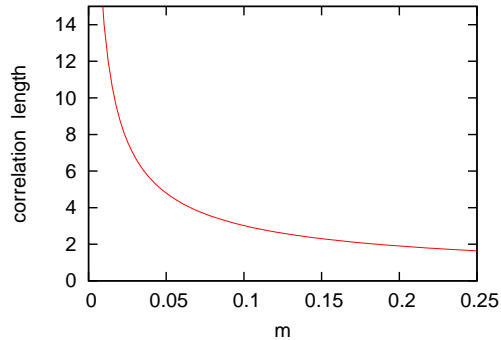}
\end{center}
\vspace*{-2mm}
\caption{The correlation length $\xi$, ranging from 
$\, \xi (m=0.24) \simeq 1.69 \,$ to $\, \xi (m=0.01) = 14.03\, $.}
\label{correl}
\end{figure}

In this way, we simulated this model at weak gauge coupling,
$\beta = 1 / g^{2} =5$, with two degenerated fermion flavours of 
mass $m= 0.01 \dots 0.24$, on $L \times L$ lattices,
$L = 16 \dots 32$ \cite{procprep}. Regarding systematic errors, 
the chiral extrapolation appears safe, and lattice spacing
artifacts are harmless as well (we always deal with smooth
configurations, plaquette values $\simeq 0.9$).
Finite size effects have to be discussed, however. To illustrate 
this, we show in Fig.\ \ref{correl} the (theoretically predicted 
\cite{Smilga}) correlation length in the regime of the fermion 
masses that we simulated.
The significance of finite size effects also implies that the
distinction between the topological sectors is important (this is
usually characteristic for the $\epsilon$-regime).
The latter are identified by measuring the fermion index \cite{Has}.
Our HMC histories contain only few topological transitions for 
the light masses, hence we performed measurements in fixed sectors.

\section{Pattern of chiral symmetry breaking}
\vspace*{-1mm}

At least in 4d Yang-Mills theory, there are only three patterns
of spontaneous chiral symmetry breaking, depending on the
fermion representation \cite{techni}
\begin{eqnarray*}
SU(N_{f}) \otimes SU(N_{f}) \to SU(N_{f}) &:& {\rm unitary} 
\hspace*{1cm} {\rm (complex~representation)} \\
SU(2N_{f}) \to O(2N_{f}) &:& {\rm orthogonal} 
\quad \, {\rm (real~representation)} \\
SU(2N_{f}) \to Sp(2N_{f}) &:& {\rm symplectic} 
\quad \ \mbox{(pseudo-real~representation)} \, . 
\end{eqnarray*}
RMT predicts the {\em unfolded level spacing
distribution} in each of these patterns \cite{HalVer}.

For a set of configurations, ${\rm conf} = 1 \dots N$,
we proceed as follows: we numerate the eigenvalues in each
configuration hierarchically, 
$\lambda_{i}^{\rm conf}$, $i= 1 \dots L^{2}$.
Now we put all eigenvalues (of all $N$ configurations)
together and order them again
hierarchically, so we attach labels $k = 1 \dots NL^{2}$.
The normalised difference 
$[k (\lambda_{i+1}^{\rm conf}) - k (\lambda_{i}^{\rm conf})] / N$
is the unfolded level spacing.
\begin{figure}[h!]
\vspace*{-2.5mm}  
\begin{center}    
\includegraphics[angle=270,width=.75\linewidth]{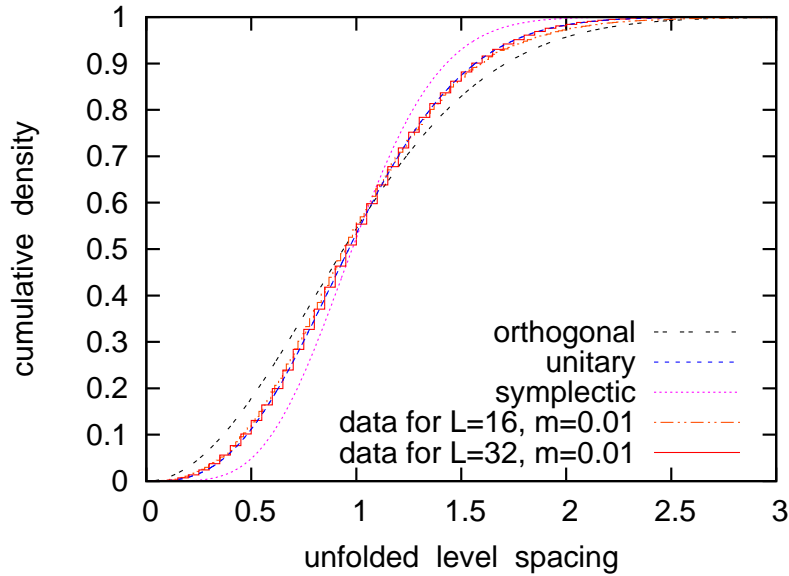}
\end{center}
\vspace*{-2mm}                                                         
\caption{The cumulative unfolded level spacing density according
to RMT (for different patterns of chiral symmetry breaking), and
from our simulation data. As in QCD, we find accurate agreement with 
the RMT curve for the unitary ensemble. A tiny deviation at 
$L=16$ disappears as the volume is enlarged.}
\label{unfoldfig}
\end{figure}

We consider the eigenvalues of $D_{\rm ovHF}^{(0)}$, {\it i.e.}\
of the Dirac operator $D_{\rm ovHF}$ that we used in the simulation,
after subtracting the mass. The eigenvalues with \, Im~$\lambda_{i} >0$
\, are mapped stereographically onto $\R_{+}\,$, \
$\lambda_{i} \to | \lambda_{i} / ( 1 - \lambda_{i}/2 ) |\,$.
Fig.\ \ref{unfoldfig} shows that the resulting level 
spacing distribution is in excellent agreement with the RMT prediction
for the unitary ensemble, as it was observed before in QCD 
\cite{BJS,QCDunfold}. We conclude that this specific RMT formula is
so generic that it is not even altered by the absence of a chiral 
condensate at $m=0$.

\vspace*{-1mm}
\section{The microscopic Dirac spectrum}
\vspace*{-1mm}

In infinite volume, $V \to \infty$, 
the chiral condensate is given by the Dirac spectrum as
\be  \label{sigspec}
\Sigma = \int d \lambda \ \frac{\rho (\lambda )}{\lambda +m}
\qquad (\rho ~ : ~ {\rm eigenvalue~density}) \ .
\ee
Along with the prediction quoted in Section 2, 
$\Sigma \propto m^{1/3}$, this suggests \cite{DHNS}
\be
\rho ( \lambda \gsim 0 ) \propto \lambda^{1/3} \ , 
\ee
in contrast to the Banks-Casher plateau that one obtains in the
standard setting (with $\Sigma (m \to 0) \neq 0$). In {\em that} case,
the density for the rescaled small eigenvalues $\lambda_{i}\Sigma V$
is scale-invariant (at fixed $m \Sigma V$) \cite{LeuSmi}.
In {\em our} case, the very general relation 
$\langle \lambda_{i} \rangle \propto [ V \rho ( \lambda \gsim 0 )]^{-1}$
implies that the parameter \cite{Poul}
\be
\zeta_{i} = \lambda_{i} V^{3/4} W_{\zeta} \qquad
({\rm for~small~} \lambda_{i})
\ee
is expected to adapt this r\^{o}le, at fixed 
$\mu_{\zeta} = m V^{3/4}  W_{\zeta}$ --- or simply at small $m$.
$W_{\zeta}$ is a constant of dimension [mass]$^{1/2}$,
which is (in this context) analogous to $\Sigma$ in the standard 
setting.

\begin{figure}[h!]
\vspace*{-1mm}
\begin{center} 
\includegraphics[angle=270,width=.47\linewidth]{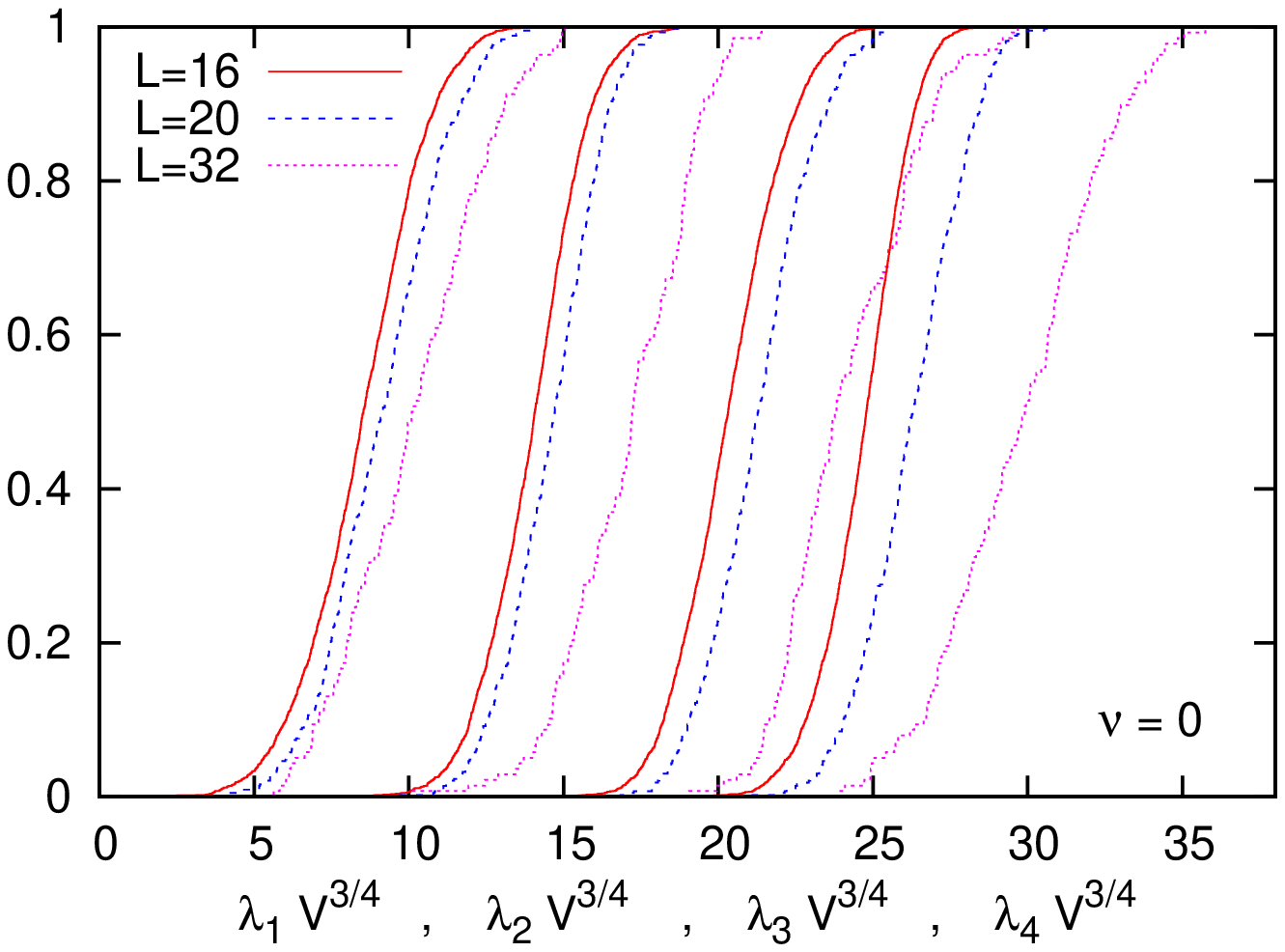}
\includegraphics[angle=270,width=.47\linewidth]{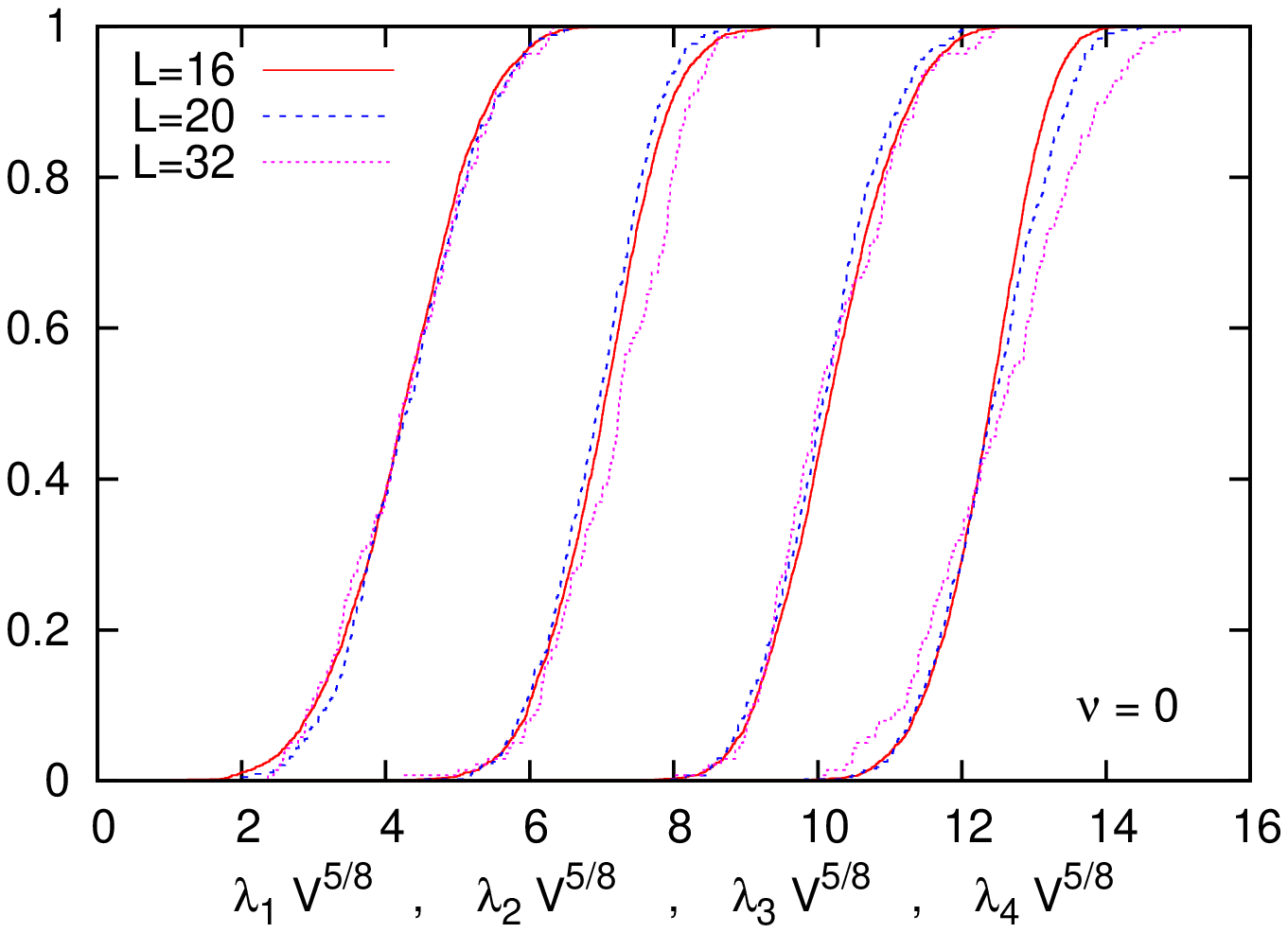}
\\
\includegraphics[angle=270,width=.47\linewidth]{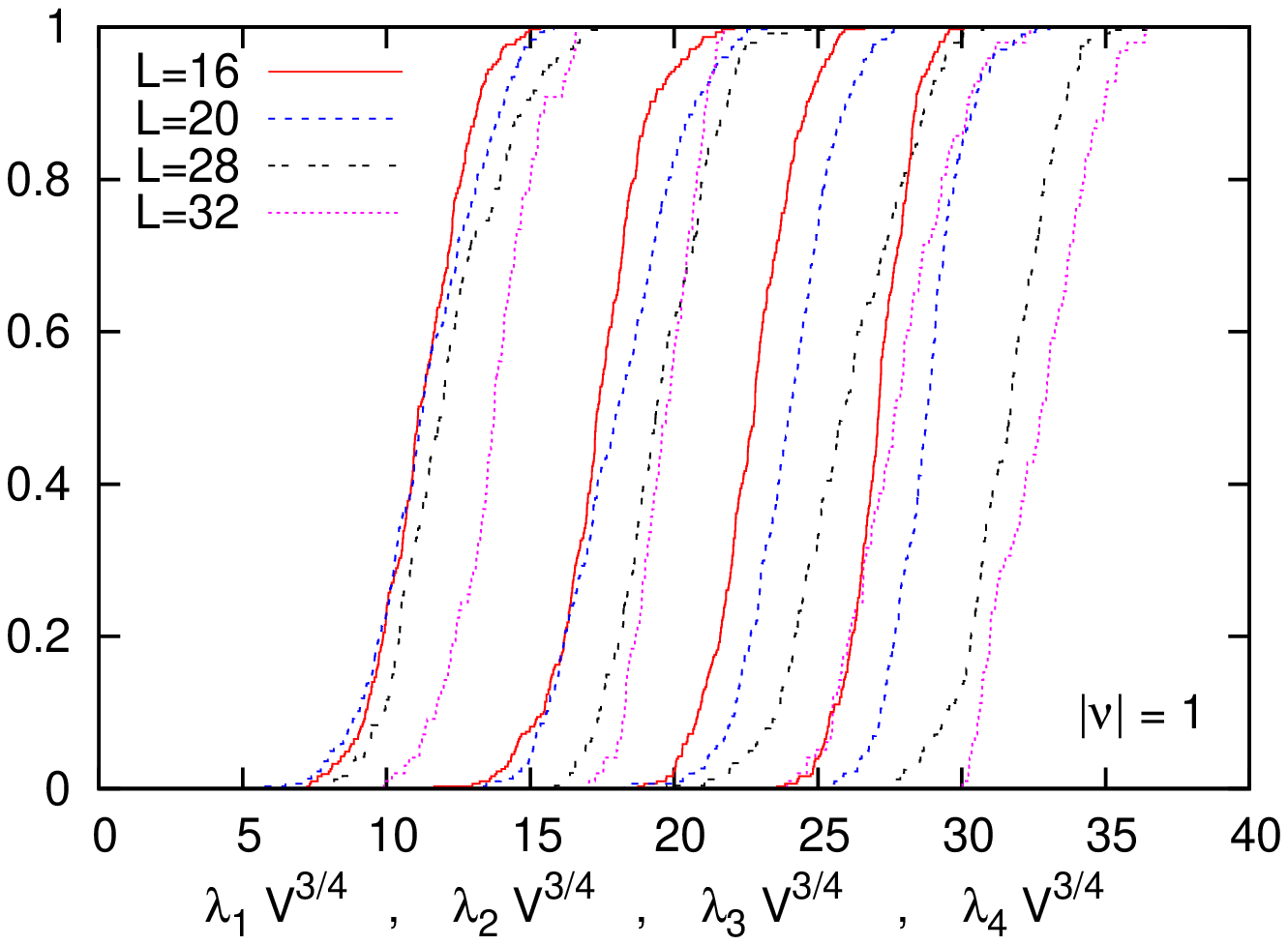}
\includegraphics[angle=270,width=.47\linewidth]{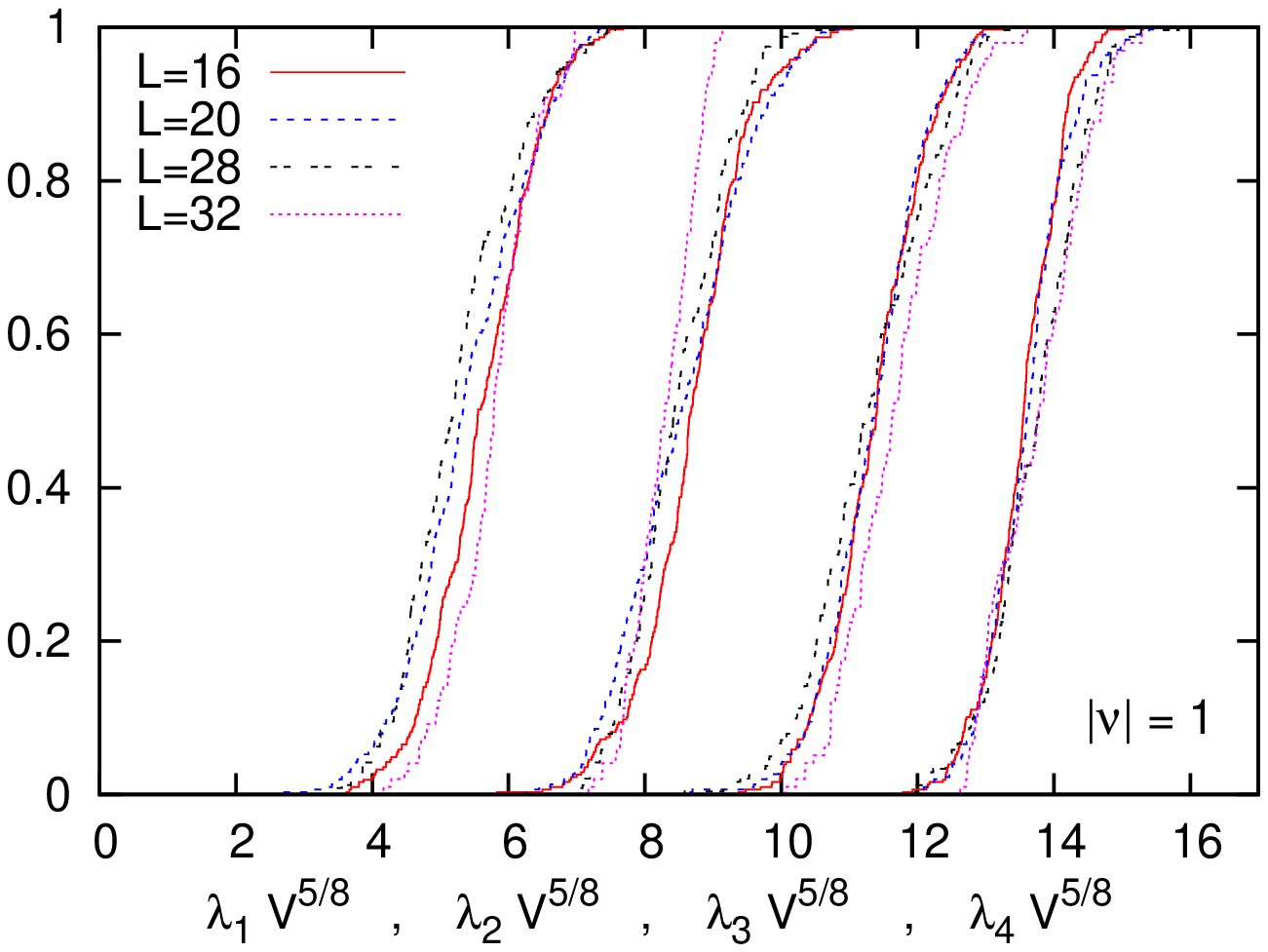}
\end{center}
\vspace*{-3mm}
\caption{Cumulative densities of $\lambda_{i} V^{3/4} \propto \zeta_{i}$ 
(left), and of $\lambda_{i} V^{5/8} \propto z_{i}$
(right), for $i=1 \dots 4$, at mass $m=0.01$ and topological charge 
$\nu =0$ (above) resp.\ $|\nu | = 1$ (below). We see that $\zeta_{i}$
strongly deviates from scale-invariance, whereas $z_{i}$ obeys this
property to an impressive precision.}
\label{lam1}
\end{figure}

\begin{figure}[h!]
\vspace*{-3mm}
\begin{center} 
\includegraphics[angle=270,width=.47\linewidth]{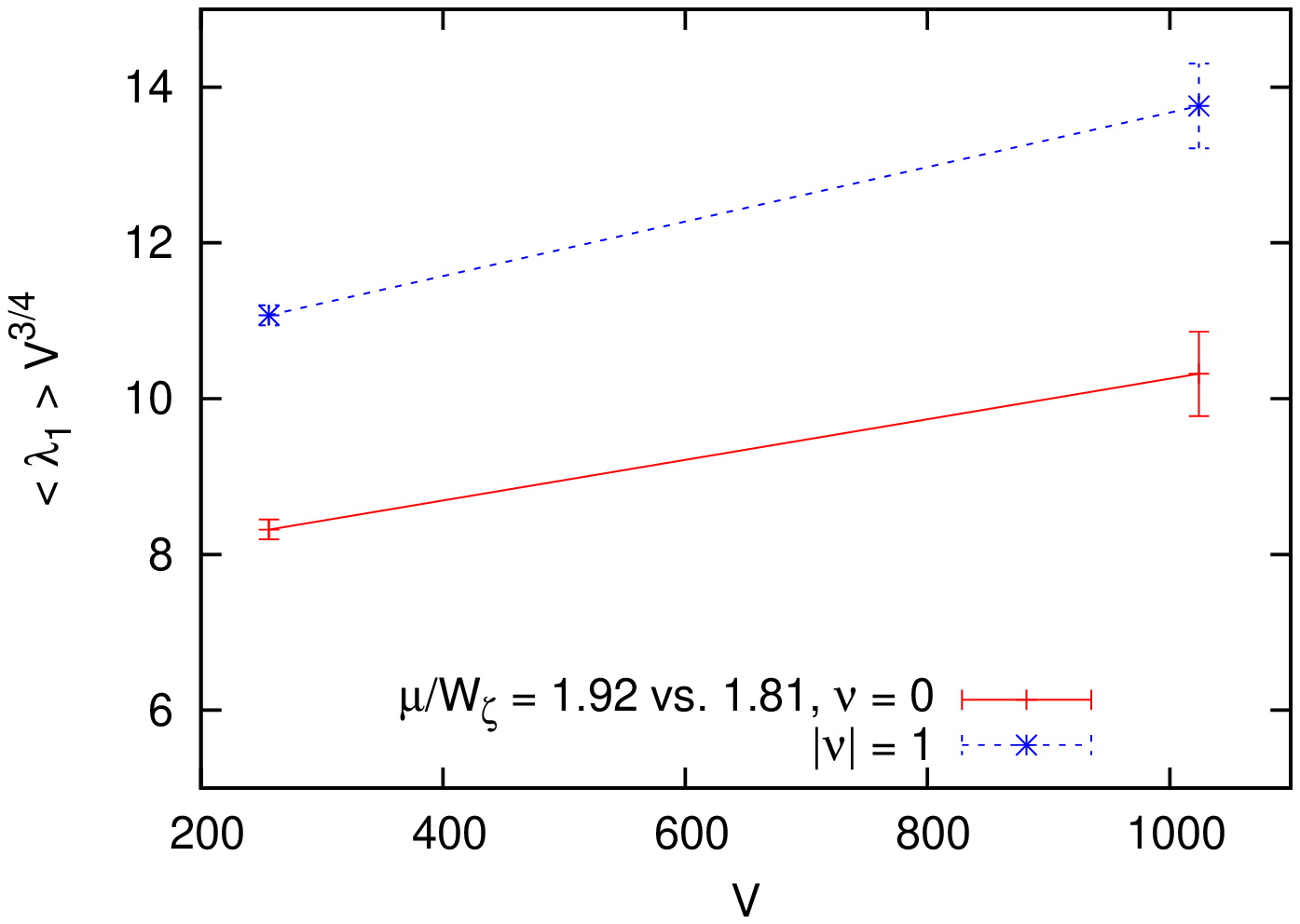}
\includegraphics[angle=270,width=.47\linewidth]{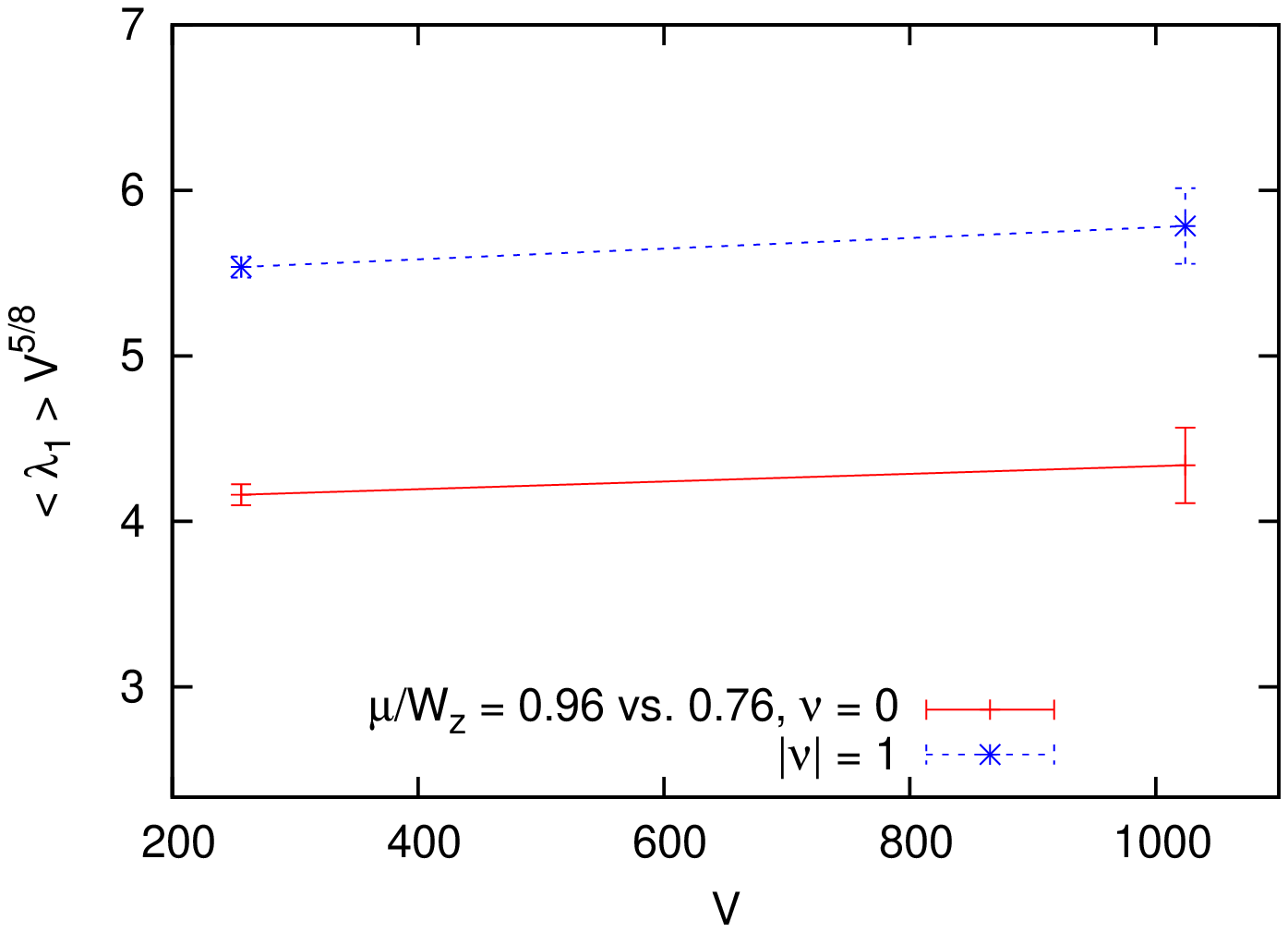}
\end{center}
\vspace*{-4mm}
\caption{Finite size scaling for $\langle \zeta_{1} \rangle$ 
at $\mu_{\zeta} \propto mV^{3/4} \approx$~const. (left) and 
$\langle z_{1} \rangle$ at $\mu_{z} \propto mV^{5/8} \approx$~const.
(right). These plots confirm again that $z_{1}$ performs
much better as a scale-invariant variable.}
\label{lam1m}
\vspace*{-3mm}
\end{figure}

Hence we probed the corresponding finite-size scaling, but it
is {\em not} confirmed. Instead our data are in excellent
agreement with a scale-invariant parameter
\be
z_{i} = \lambda_{i} V^{5/8} W_{z} \qquad
(W_{z} ~ : ~ {\rm constant~of~dimension~[mass]^{1/4}}) \ .
\ee
This is illustrated in Fig.\ \ref{lam1} for our lightest fermion mass,
$m=0.01$, in the sectors of topological charge $\nu =0$ and $|\nu |= 1$.
We also tested the behaviour if the rescaled mass
is kept $\approx$~const., as an alternative to just keeping
$m$ very small. In Fig.\ \ref{lam1m} we compare 
$\langle \zeta_{i} \rangle$ in different volumes, 
$V = L^{2}, \ L=16$ and $32$,
again in the sectors $| \nu | =0$ and $ 1$, for 
$\mu_{\zeta} \approx$~const. We add the corresponding test
with $\langle z_{i} \rangle$ and $\mu_{z} = m V^{5/8}  W_{z} 
\approx$~const., which displays again a superior finite size 
scaling.

\begin{figure}[h!]
\vspace*{-3mm}
\begin{center} 
\includegraphics[angle=270,width=.46\linewidth]{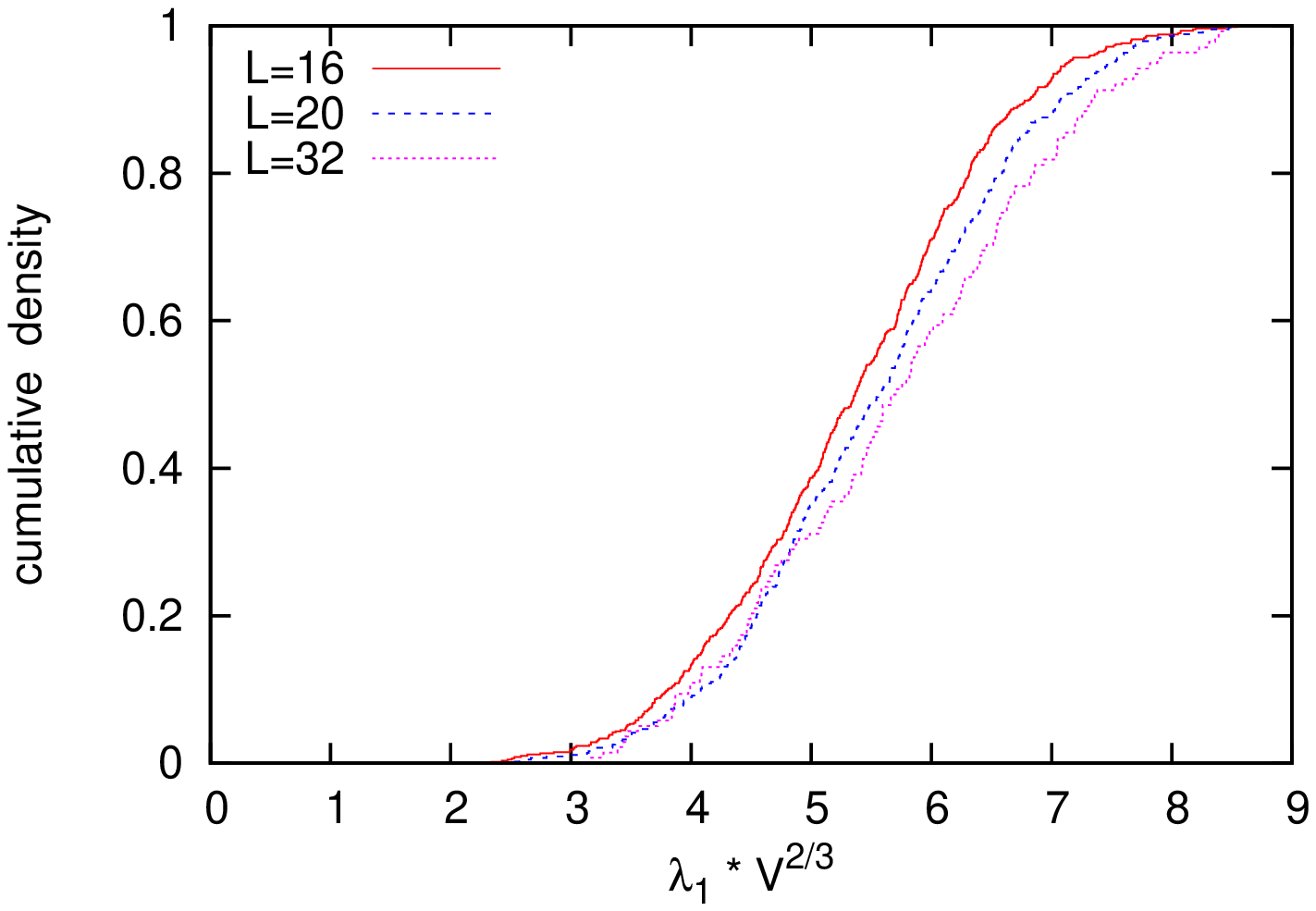}
\includegraphics[angle=270,width=.46\linewidth]{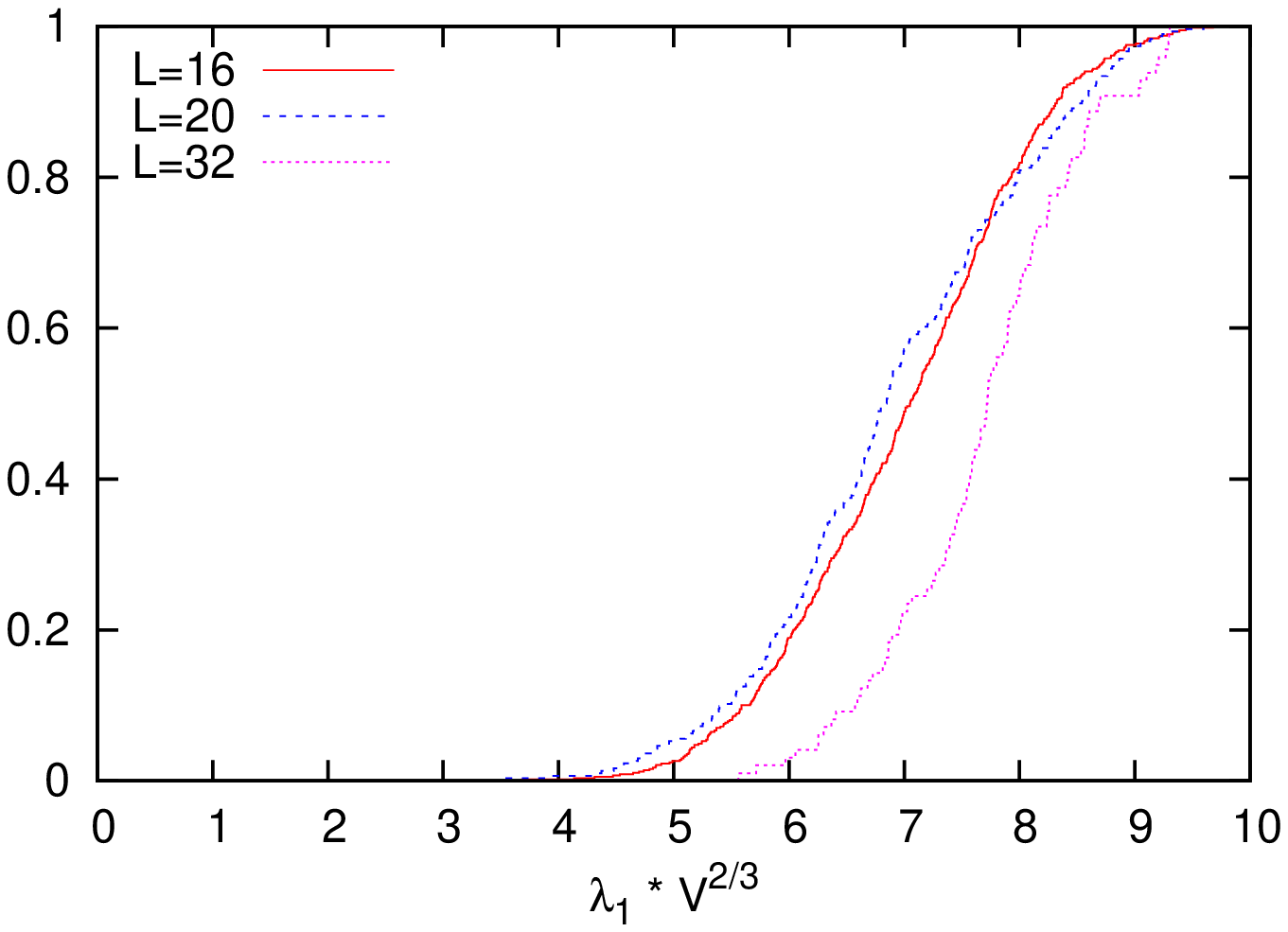}
\vspace*{-3mm} \\
\includegraphics[angle=270,width=.5\linewidth]{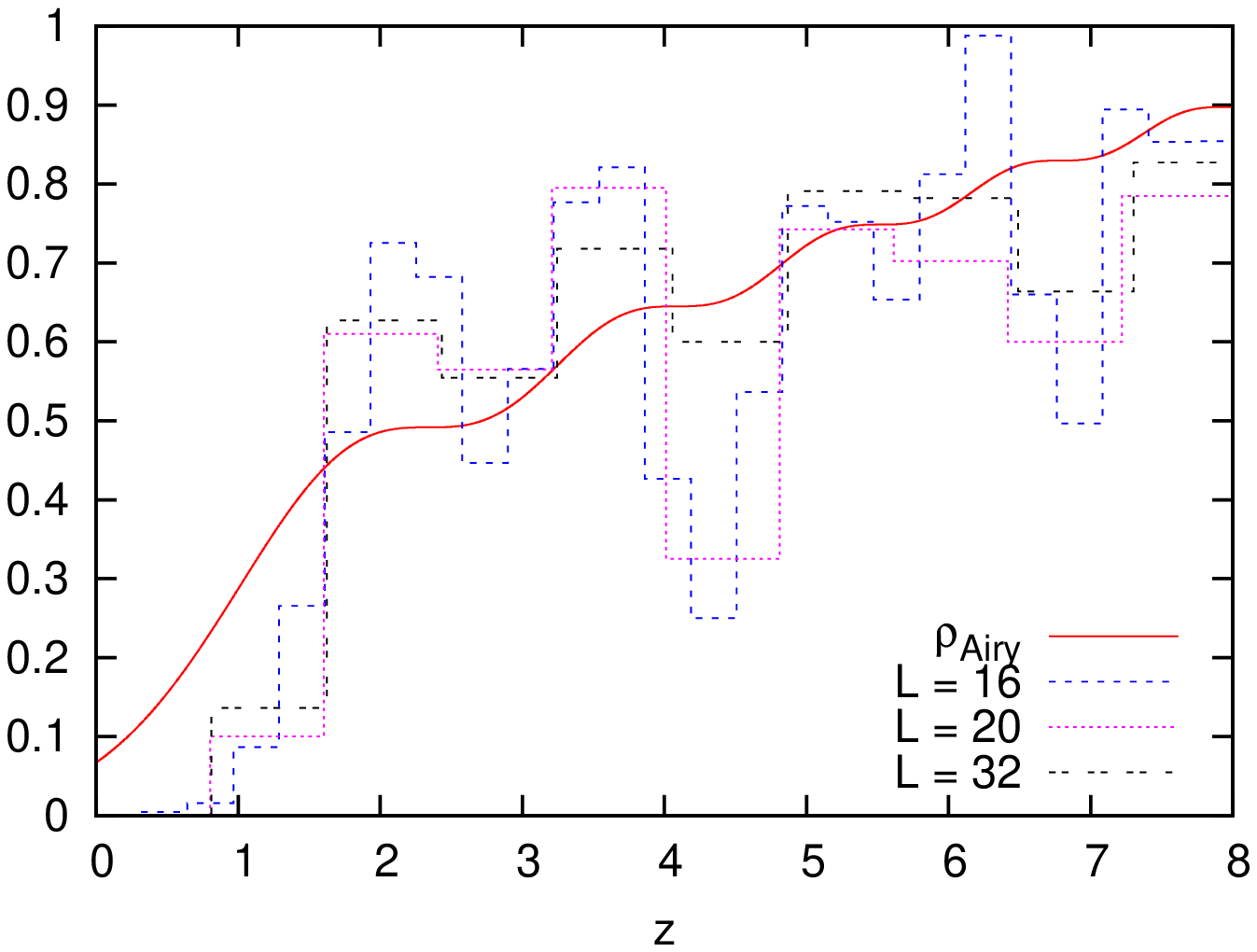}
\end{center}
\vspace*{-4mm}
\caption{{\it Above:} Finite size scaling for $Z_{1}
\propto \lambda_{1} V^{2/3}$, at $m=0.01$ and
$\nu = 0$ (left), or $|\nu |= 1$ (right). Regarding scale-invariance,
$Z_{1}$ performs better than $\zeta_{1}$, but not as good as $z_{1}$.
{\it Below:} Eigenvalue histogram for $m=0.01$, $\nu =0$ compared to the
spectral density $\rho_{\rm Airy}$ in eq.\ (5.5),
which RMT predicts in this case.
None of these three plots does convincingly support this scenario, in contrast
to the compelling evidence that we found for that finite-size scaling of
the variable $\, z_{i} = \lambda_{i} V^{3/5} W_{z}\,$.}
\label{lam1Z}
\vspace*{-2mm}
\end{figure}

To complete this discussion, we consider even a third scenario,
where the exponent of $V$ in the rescaling factor is between
the two cases considered so far: now the scale-invariant variable
would be $Z_{i} = \lambda_{i} V^{2/3} W_{Z}$ ($W_{Z}$ of dimension
[mass]$^{1/3}$). The behaviour of $Z_{1}$ is
shown in Fig.\ \ref{lam1Z} (plots above); 
as in Fig.\ \ref{lam1} we fix $m=0.01$ and consider 
$|\nu | = 0$ and $1$. The finite size scaling quality is clearly 
better than the one of $\zeta_{1}$, but it cannot compete with $z_{1}$.
This third scenario belongs to a theoretically
explored universality class: it corresponds to 
$\rho ( \lambda \gsim 0) \propto \lambda^{1/2}$, which is the spectral
density obtained by RMT in the {\em Gaussian approximation}. There
is a detailed prediction for the spectral density in terms of {\em
Airy functions} Ai \cite{airyref},
\be \label{airyfun}
\rho_{\rm Airy} (Z) \propto Z [ {\rm Ai}(-Z)]^{2} + [ {\rm Ai}'(-Z)]^{2} 
\quad (\sim \sqrt{Z} /\pi \quad {\rm at~}Z \gg 1) \ .
\ee
Fig.\ \ref{lam1Z} (below) compares this function to the histogram
that we obtained in various volumes at $m=0.01$ and $\nu =0$.
Our data display a far more marked wiggle structure, hence this
agreement is not convincing, and we stay with
$z_{i} = \lambda_{i} V^{5/8} W_{z}$ as the clearly preferred 
scale-invariant variable --- even though no theoretical prediction
for the detailed structure of $\rho (z)$ has been worked out so far.
The error on the exponent $5/8$ will be estimated later,
see last entry in Ref.\ \cite{procprep}.

\vspace{-3mm}
\section{Higher eigenvalues}
\vspace{-2mm}

\begin{figure}[h!]
\vspace*{-1mm}
\begin{center} 
\includegraphics[angle=270,width=.47\linewidth]{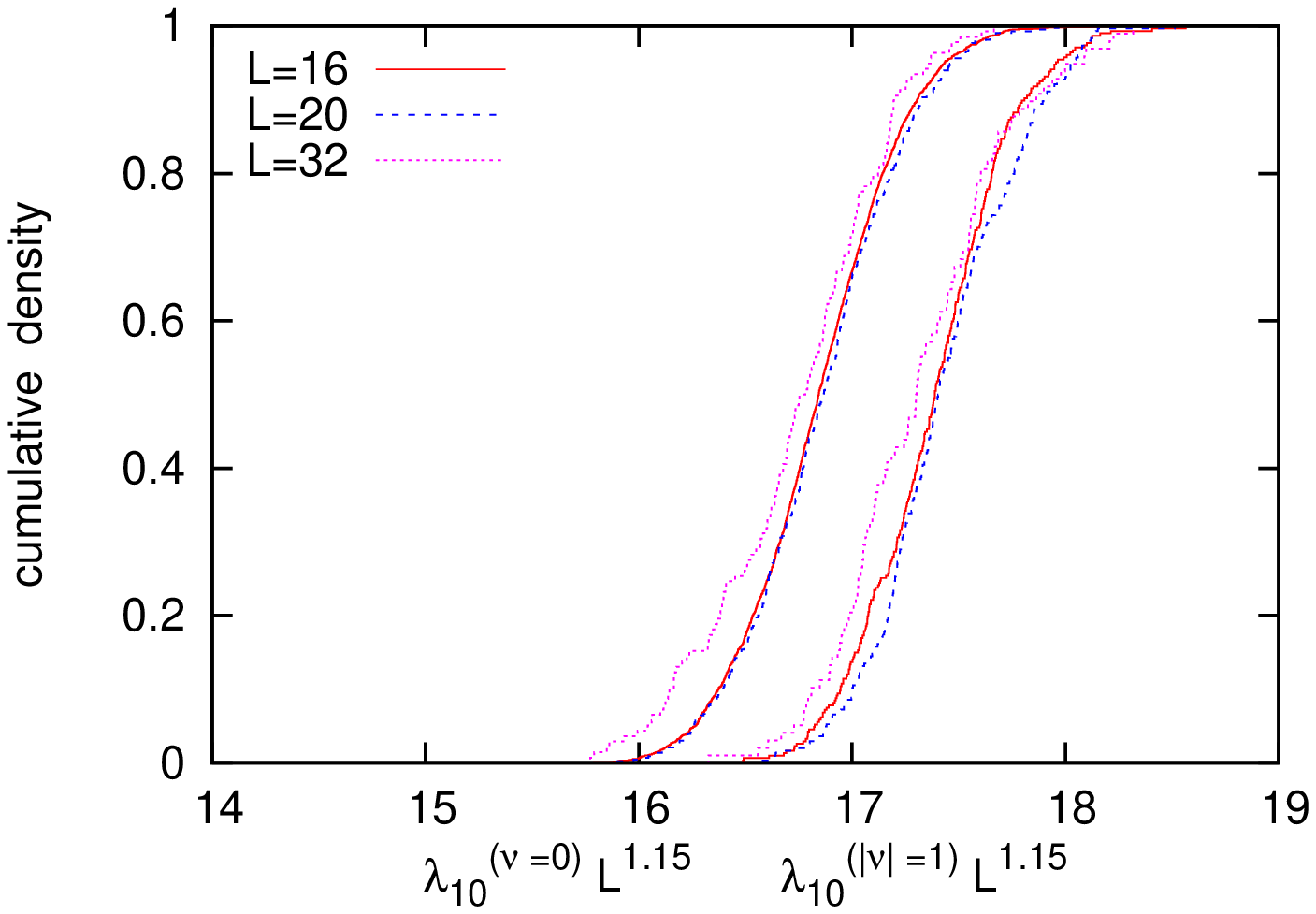}
\includegraphics[angle=270,width=.47\linewidth]{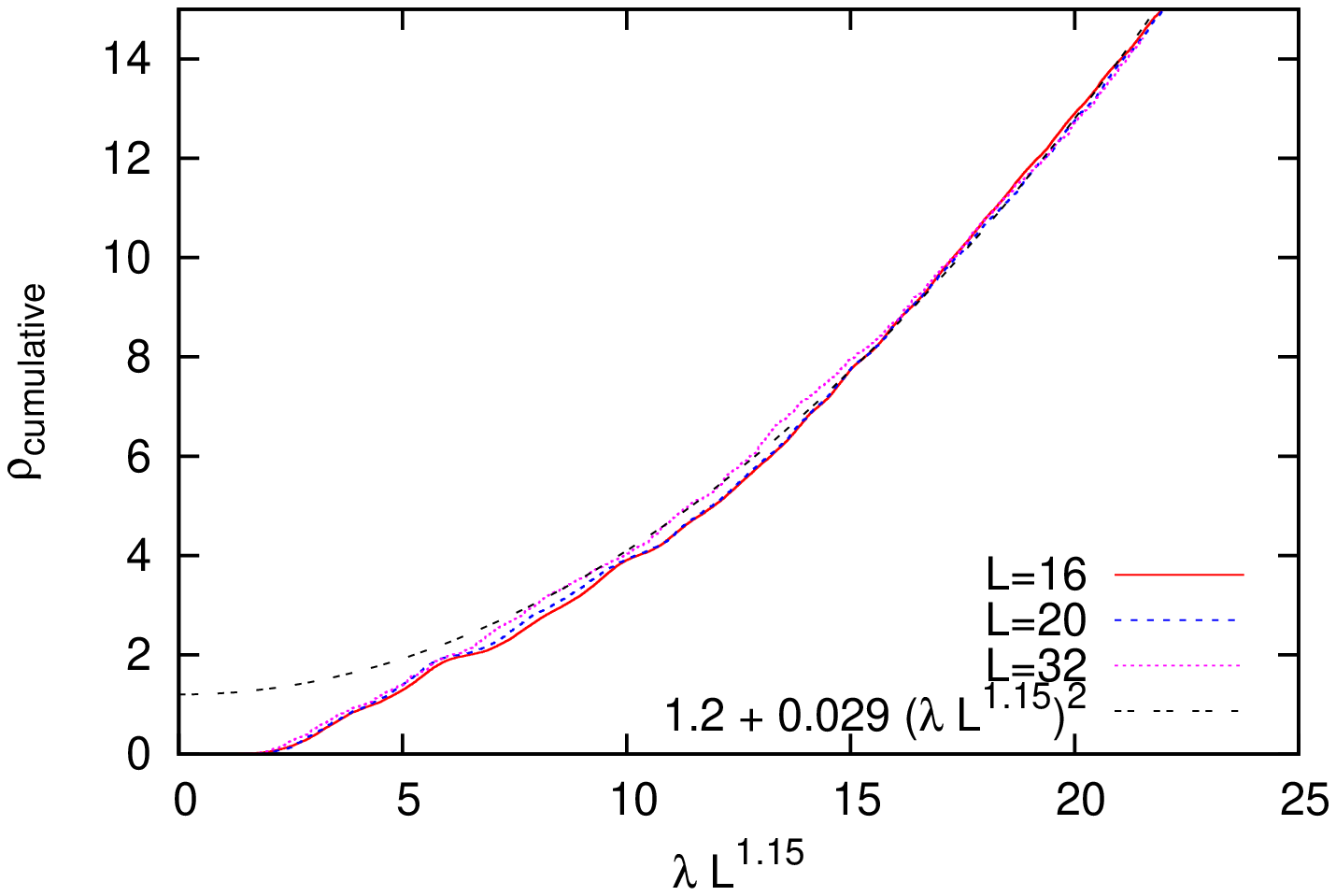}
\end{center}
\vspace*{-3mm}
\caption{For $\lambda_{10}$, a bulk eigenvalue, the scaling
factor is shifted to $L^{1.15}$. The rescaled full spectral
cumulative densities in different volumes (for $m=0.01$, $\nu = 0$) 
agree well, and turn into bulk behaviour $\rho (\lambda ) 
\propto \lambda$, (resp.\ $\rho_{\rm cumulative} (\lambda ) 
\propto \lambda^{2}$), which is expected in $d=2$.}
\label{lam10}
\end{figure}

At last we take a look at $\lambda_{10}$ as one of the bulk
eigenvalues, and we find a optimal finite size scaling for
$\lambda_{10} L^{1.15}$, see Fig.\ \ref{lam10} (left). 
The plot on the right shows that
this factor works well also for the rescaled full cumulative 
density (including all eigenvalues up to the considered value).
Based on the fact that the spectral cutoff $\lambda_{\rm max}=2$ 
is fixed in any volume, it is now tempting to speculate that the
volume factor for a good finite size scaling gradually decreases
from $V^{3/5} \dots V^{0}$. However, considering eigenvalues
above the regime shown in Fig.\ \ref{lam10}, but below the
cutoff regime, we could not find any consistent scaling 
factor --- and indeed there is no need for it to exist.

\vspace{-3mm}
\section{Conclusions}
\vspace{-2mm}

We presented a pioneering numerical study of a microscopic
Dirac spectrum near a chiral limit with $\Sigma (m \to 0) =0$
at zero temperature.
In particular we analysed spectral data of the $N_{f}=2$ 
Schwinger model, obtained from simulations with dynamical
chiral fermions.

The unfolded level spacing density follows the RMT formula for the 
unitary ensemble.

Refs.\ \cite{Smi92,Smilga} predict $\Sigma (m) \propto m^{1/3}$, which
suggests a microscopic density $\rho (\lambda \gsim 0) \propto 
\lambda^{1/3}$, and the scale-invariant variable 
$\zeta \propto \lambda V^{3/4}$
(generally $\rho \propto \lambda^{\alpha}$ suggests  
a scale-invariant $\propto \lambda V^{1/(1+\alpha)}$).
However, this conjecture does not agree with our data.
(Note that its derivation may be invalidated by inserting a
spectral density with explicit mass-dependence, 
$\rho (\lambda, m)$, in eq.\ (\ref{sigspec}).)

An alternative scenario with 
$\rho (\lambda \gsim 0) \propto \lambda^{1/2}$ is favoured compared to
the initial guess. It has a known theoretical background, but the
data do not strongly support it either.

Our data strongly favour $z \propto \lambda V^{5/8}$
as the scale-invariant variable, and therefore 
$\rho (\lambda \gsim 0) \propto \lambda^{3/5}$.
This determination is much more reliable than
a direct fit to the measured $\rho (\lambda)$. 
Interestingly, the first work in Ref.\ \cite{Smilga} specifies 
that $\Sigma \propto m^{1/3}$ is expected for 
$\ell = \sqrt{2} m L^{3/2}/(\beta \pi)^{1/4} \gg 1$, whereas
$\ell \ll 1 \ll 2L / \sqrt{\pi \beta}$ implies $\Sigma \propto mL$. 
For $m=0.01$ we are in an {\em intermediate} regime, $\ell = 0.5 \dots
1.3$ \ (and \ $2L / \sqrt{\pi \beta} = 8.1 \dots 16.2$), which 
renders our exponent in $\Sigma \propto m^{3/5}$ plausible.

For a precise theoretical test,
we hope for RMT formulae to be worked out for this setting,
so they can be confronted with our results; this is not 
straightforward, but it may be feasible \cite{Poul}. \\

\vspace*{-3mm}
\noindent
{\bf Acknowledgements :} {\it We are indebted to Stanislav Shcheredin and
Jan Volkholz for their contributions to this work at an early stage,   
and to Poul Damgaard, Hidenori Fukaya and Jim Hetrick  
for numerous highly enlightening and helpful discussions.}


\vspace*{-3mm}
\begin{thebibliography}{99}
\vspace*{-1mm}

\bibitem{stag} For a review, see \ J.J.M.\ Verbaarschot and T.\ Wettig,
{\em Ann.\ Rev.\ Nucl.\ Part.\ Sci.}\ {\bf 50} (2000) 343.

\bibitem{BJS} W.~Bietenholz, K.~Jansen and S.~Shcheredin, 
\emph{JHEP} {\bf 07} (2003) 033. 

\bibitem{GWRMT} L.~Giusti, M.~L\"uscher, P.~Weisz and H.~Wittig, 
\emph{JHEP} {\bf 11} (2003) 023. 
H.\ Fukaya {\it et al.} (JLQCD Collaboration),
{\em Phys.\ Rev.\ Lett.}\ {\bf 98} (2007) 172001.


\bibitem{airyref} P.H.\ Damgaard, U.M.\ Heller, R.\ Niclasen,
and K.\ Rummukainen, {\em Nucl.\ Phys.} {\bf B 583} (2000) 347.

\bibitem{highT} F.\ Farchioni, P.\ de Forcrand, I.\ Hip,
C.B.\ Lang and K.\ Splittorff, {\em Phys.\ Rev.} {\bf D 62} 
(2000) 014503.
T.\ Kov\'{a}cs, arXiv:0906.5373 [hep-lat].

\bibitem{procprep} J.\ Volkholz, W.\ Bietenholz and S.\ Shcheredin, 
{\em PoS(LAT2006)040}. 
W.\ Bietenholz, S.\ Shcheredin and J.\ Volkholz,
{\em PoS(LAT2007)064}. W.\ Bietenholz and I.\ Hip,
{\em PoS(LAT2008)079}. 
W.\ Bietenholz, I.\ Hip, S.\ Shcheredin and J.\ Volkholz, in prepration.

\bibitem{Smi92} A.\ Smilga, {\em Phys.\ Lett.}\ {\bf B 278} (1992) 371.
Y.\ Hosotani and R.\ Rodriguez,
{\em J.\ Phys.}\ {\bf A 31} (1998) 9925.

\bibitem{DHNS} P.H.\ Damgaard, U.M.\ Heller, R.\ Narayanan
and B.\ Svetitsky,
{\em Phys.\ Rev.} {\bf D 71} (2005) 114503. 

\bibitem{CJS} S.R.\ Coleman, R.\ Jackiw and L.\ Susskind,
{\em Annals Phys.} {\bf 93} (1975) 267.

\bibitem{Smilga} J.\ Hetrick, Y.\ Hosotani and S.\ Iso,
{\em Phys.\ Lett.}\ {\bf B 350} (1995) 92.
A.\ Smilga, {\em Phys.\ Rev.}\ {\bf D 55} (1997) 443.

\bibitem{Hide} H.\ Fukaya and T.\ Onogi,
{\em Phys.\ Rev.}\ {\bf D 70} (2004) 054508. 

\bibitem{WBIH} W.\ Bietenholz and I.\ Hip, {\em Nucl.\ Phys.}\ 
{\bf B 570} (2000) 423.

\bibitem{WBEPJC} W.\ Bietenholz, {\em Eur.\ Phys.\ J.} {\bf C 6} (1999) 537.

\bibitem{Neu} 
H.\ Neuberger, {\em Phys.\ Lett.} {\bf B 417} (1998) 141.

\bibitem{ML} M.\ L\"{u}scher, 
\emph{Phys.\ Lett.} {\bf B 428} (1998) 342. 

\bibitem{DovHF} W.\ Bietenholz, {\em Nucl.\ Phys.} {\bf B 644} (2002) 
223. S.\ Shcheredin, Ph.D.\ Thesis (Humboldt Univ.\ Berlin) [hep-lat/0502001].
W.\ Bietenholz and S.\ Shcheredin, {\em Nucl.\ Phys.} 
{\bf B 754} (2006) 17.

\bibitem{Has} P.~Hasenfratz, V.~Laliena and F.~Niedermayer, 
{\it Phys.\ Lett.} {\bf B 427} (1998) 125. 

\bibitem{techni} S.\ Dimopoulos, {\em Nucl.\ Phys.} {\bf B 168} (1980) 69.
M.E.\ Peskin, {\em Nucl.\ Phys.} {\bf B 175} (1980) 197.
J.P.\ Preskill, {\em Nucl.\ Phys.} {\bf B 177} (1981) 21.

\bibitem{HalVer}  M.A.\ Halasz and J.J.M.\ Verbaarschot,
{\em Phys.\ Rev.\ Lett.} {\bf 74} (1995) 3920.

\bibitem{QCDunfold} R.G.\ Edwards, U.M.\ Heller, J.E.\ Kiskis and
R.\ Narayanan,                                              
{\em Phys.\ Rev.\ Lett.} {\bf 82} (1999) 4188. 

\bibitem{LeuSmi} H.\ Leutwyler and A.\ Smilga, 
{\em Phys.\ Rev.} {\bf D 46} (1992) 5607.

\bibitem{Poul} P.H.\ Damgaard, private communication.

\end{thebibliography}
\end{document}